\documentclass[aps,reprint,superscriptaddress,showpacs,floatfix,longbibliography]{revtex4-1}
\usepackage{graphicx,subfigure,epsfig,psfrag,verbatim,amsmath,amssymb,xcolor}
\usepackage{bbm}
\usepackage{amsopn}
\usepackage{natbib}
\usepackage[colorlinks=true,citecolor=blue,urlcolor=blue,linkcolor=blue,hyperindex]{hyperref}

\makeatletter
\input epsf

\def\be{\begin{equation}}
\def\ee{\end{equation}}
\def\ba{\begin{eqnarray}}
\def\ea{\end{eqnarray}}
\makeatother

\allowdisplaybreaks

\begin{document}
\title{Opto Propeller Effect on Chiral Micro-Rotors}
\author{Yiwen Tang and Zhibing Li$^{*}$}

\affiliation{School of Physics, Sun Yat-sen University, Guangzhou 510275, People's Republic of China}

\date{\today}

\begin{abstract}
Manipulating mega biomolecules and micro-devices with light is highly appealing. Opto driving torque can propel micro-rotors to translational motion in viscous liquid, and then separate  microsystems according to their handedness. We study the torque of dielectric loss generated by circular polarized lasers. The unwanted axial force which causes the handedness independent translational motion is cancelled by the counter propagating reflection beams. The propelling efficiency and the friction torque of water are obtained by solving the Navier-Stokes equation. In the interesting range of parameters, the numerical friction torque is found to be linear in the angular velocity with a slope depending on the radius or rotor as $r^3$. The time-dependent distribution of angular velocity is obtained as a solution of the Fokker-Planck equation, with which the thermal fluctuation is accounted. The results shed light on the micro-torque measurement and suggest a controllable micro-carrier.
\end{abstract}

\maketitle

\section{Introduction}

People have been interested in the separation of chiral micro-systems by propeller effect for many years\cite{BARANOVA1978435,doi:10.1021/ja405705x,doi:10.1021/nl900186w,doi:10.1063/1.2428249,doi:10.1063/1.2107867,clemens2015molecular}. Various optical techniques have been implemented for rotating or propelling mega biomolecules\cite{doi:10.1146/annurev-biophys-083012-130412} or microelectromechanical systems\cite{doi:10.1063/1.1339995,doi:10.1080/09500349514550171,PhysRevLett.75.826} and for direct measurement of micro-torque. It has been known that the rotating electric field can induce a constant torque to a polarizable conductive particle\cite{BERG19932201,216534}, whose dipole rotates at the same frequency as the electric field but not synchronously due to the conductance or dielectric loss. The driving torque can be generated by transferring of either spin or orbital angular momentum of light to matter\cite{simpson1997mechanical}. Early demonstrations depend on the breaking rotational symmetry of the particle or the trapping beam\cite{PhysRevA.68.033802,friese1998optical,galajda2003orientation,o2002rotational}. Friese et.al have shown that both linearly and circularly polarized light can rotate a microscopic birefringent particle\cite{friese1998optical}. It has been shown that the angular optical trap exerts torque on individual biological molecules\cite{deufel2007nanofabricated,PhysRevLett.92.190801}.

Recently, the propelling effect of micro-systems rotated by light has been studied quantitatively with specific model parameters\cite{doi:10.1063/1.4962411,PhysRevFluids.2.064303}. They prove theoretically that the separation of chiral micro-systems by this mean is feasible even when the thermal fluctuation is taken into account. A designed nano-turbine driven by fluid flow has been simulated with molecular dynamics\cite{li2014rotation}. They found that the rotational angular velocity has a robust linear relationship with the fluid flow velocity. The ratio of the angular velocity to the translational velocity is much smaller than the ideal value, implying a large edge effect on propelling efficiency for the nano-turbine. A natural question is: how small is a system whence its edge effect becomes important?  For applications, people also concern over the feasibility and the efficiency of optical separation of chiral microsystems or mega biological particles and over how they depend on the structure of the system and the temperature. To answer these questions quantitative calculations which depend on the system parameters are required.

We carry out a qualitative calculation/simulation  for the propeller effect of a special type of chiral micro-systems driven by rotating electric fields. Our target is very similar to a recent work by Makino and Doi\cite{PhysRevFluids.2.064303}. But unlike their particle that has a permanent electric dipole, we consider the chiral microsystems that are made in a dielectric material without permanent electric dipole. We investigate both rotational and translational motions of a micro-rotor immersed in water, simulating a micro-biological system or micro-machine in living/working environment. The micro-rotor is driven by circularly polarized laser beams. We are interested in the opto torque originated from the dielectric loss that leads to handedness-depending translational motion through the propeller effect. In contrast, the translational motion caused by direct momentum transfer in photon scattering do not depend on the handedness of the rotor. To cancel the force and torque from direct momentum transfer, we propose to use two laser beams, both of them propagate along the axis but in opposite directions.

The hydrodynamics of the  micro-rotor in water is described by the Navier-Stokes equation. In principle it gives the propelling efficiency and the friction torque, which are two crucial quantities in the present text. However, a general solution for our model is still laking because of the significant edge effect of the micro-rotor. Therefore we will partially resort to numerical simulation. Numerical friction torque exhibits a cubic scaling law in the regime of small Reynolds number, that is consistent with the dimension analysis. The cubic scaling law is broken at a radius of hundred micrometers, revealing the crossover from the micro-sized regime to the macro-sized regime where the edge effect is less important and the fluid on the blade surface can be described by the model of laminar boundary layer.

Moreover, the thermal fluctuation that competes with the hydrodynamic motion is also significant for small Reynolds number. It will be accounted by random torques. Langevin-like stochastic equation of motion is converted into the Fokker-Planck equation. The later is solved to obtain the time-depending distribution of angular velocity. Combining the propelling efficiency from the simulation, we obtain the time-dependent translational velocity for a representative micro-rotor made with graphite and carbon microtube.

In the following, the model and the theoretical framework are described in Sec. \uppercase\expandafter{\romannumeral2}. In Sec. \uppercase\expandafter{\romannumeral3} the opto driving torque is derived and the scaling behavior of the friction torque is discussed based on a general dimension analysis as well. Sec. \uppercase\expandafter{\romannumeral4} gives the solution of the Fokker-Planck equation for time-depending distribution of the angular velocity in the regime of small Reynolds number. Sec. \uppercase\expandafter{\romannumeral5} describes the numerical solution of the Navier-Stokers equation. The translational velocity is obtained for specific parameters of the material and laser in Sec. \uppercase\expandafter{\romannumeral6}. Summary is given in the last section.

\section{model and theory}

\begin{figure}
  \centering
  \subfigure[]{
  \includegraphics[width=8.6cm]{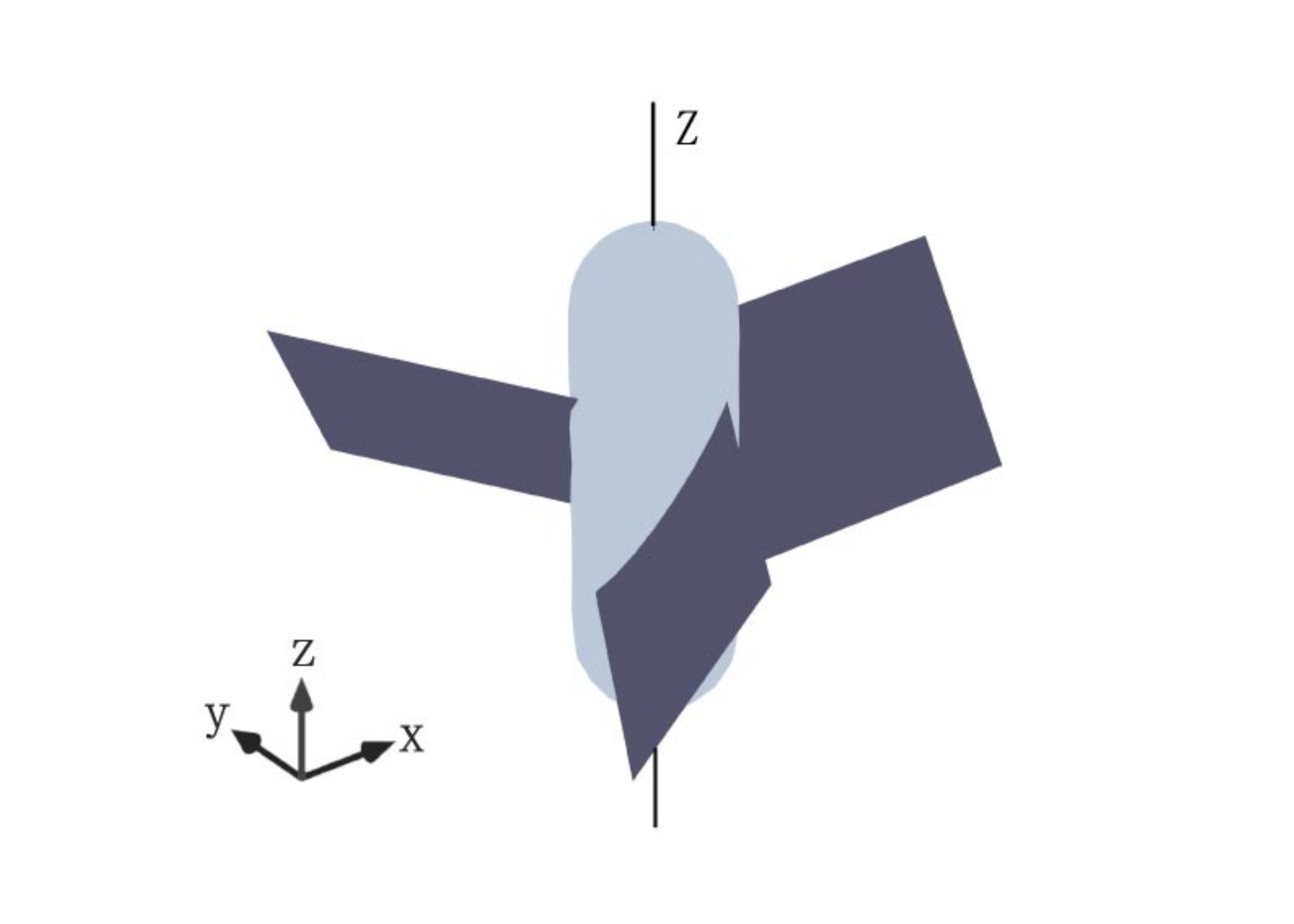}}
  \subfigure[]{
  \includegraphics[width=3.2cm]{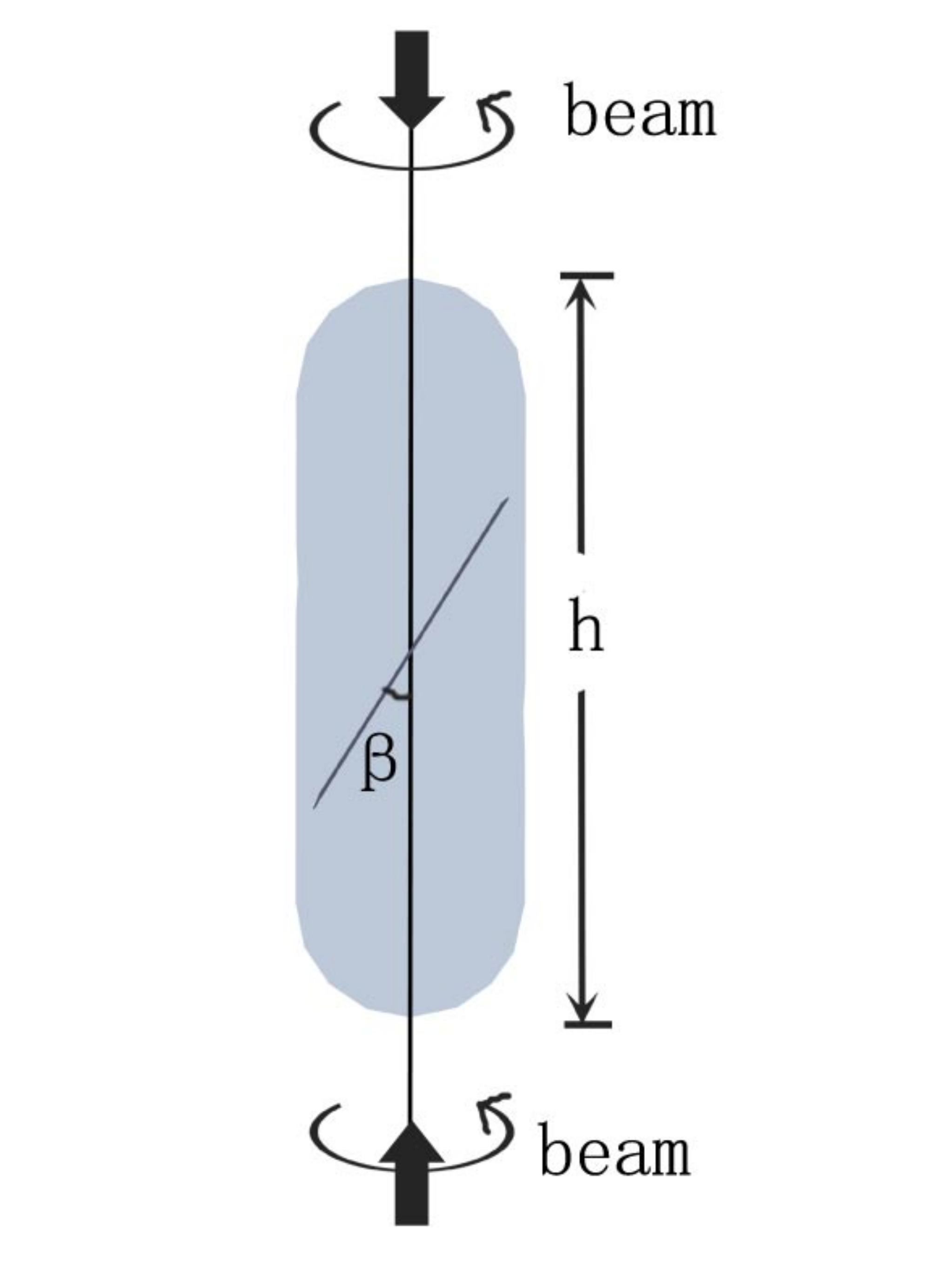}}
  \subfigure[]{
  \includegraphics[width=5cm]{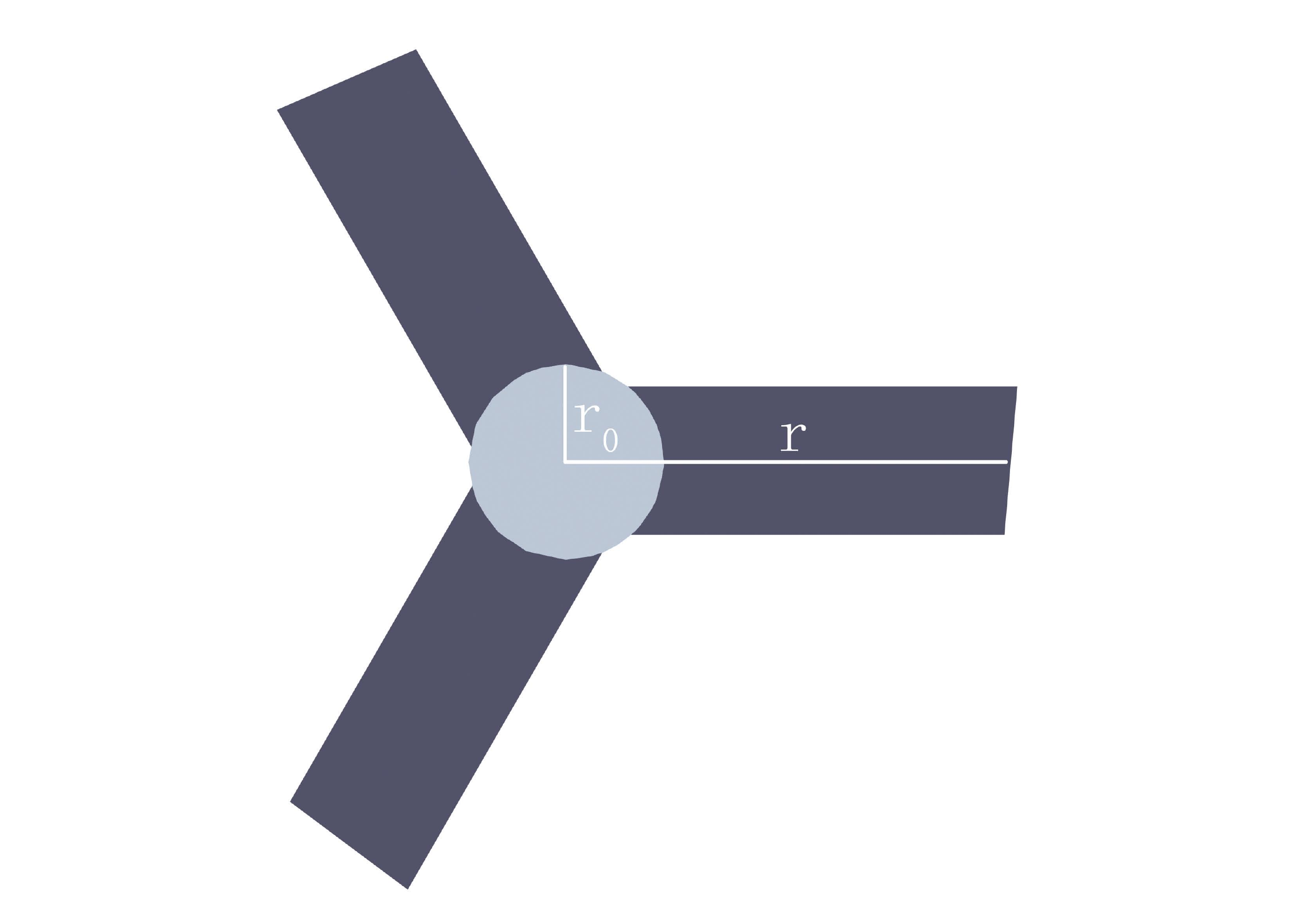}}\\
  \caption{(a) The left-handed micro-rotor with three symmetric graphite blades. (b) The side view of the micro-rotor. The blades are combined with the CMT at a tilt angle $\beta$. Propagating directions of two laser beams are indicated by two bold arrows and the polarizations are indicated by two return arrows, on the bottom and top respectively. (c) The top view. Each blade has radius $r$ and area $S$. }\label{Fig.turbine}
\end{figure}

We consider a micro-rotor that consists of a carbon microtube (CMT) \cite{WEN20114067} and three symmetric graphite blades, as shown in Fig.\ref{Fig.turbine}. The capped CMT has radius $r_0$  and height $h$ along the axle of the rotor, the $z$-axis; while three blades, each of which has area $S$ and thickness $a$, are combined with the CMT at a tilt angle $\beta$. All lengths are scaling with the rotor radius $r$. In other words, their ratios with $r$ are fixed.

The micro-rotor is driven by two counter propagating circularly polarized laser beams as shown in Fig.{\ref{Fig.turbine}(b). The reason to use two beams instead of one is to cancel the force and torque due to direct momentum transfer in photon scattering. They are not wanted because they drive both right-handed and left-handed rotors to move in the same direction along the axis.
Since a circularly polarized beam has spin angular momentum, there is also angular momentum transferring that is expressed as dielectric loss of electric field in blades in classical electrodynamics. As a consequence, the blades are exerted by a torque from the dielectric loss. The direction of this torque is fixed by the rotation direction of the electric field. Therefore the propeller thrusts originated from the dielectric loss of a circularly polarized beam have opposite directions for the right-handed and left-handed rotors, enabling the  chiral separation. In  Fig.\ref{Fig.turbine}(b), both the upward and downward propagating laser beams have electric fields rotating in the same direction thus create the same torque by dielectric loss.

Denoting the opto driving torque due to the dielectric loss as $Q^p$ and the friction torque of the viscous liquid as $Q^f$, which will be discussed in Sec. \uppercase\expandafter{\romannumeral3}, and the random torque as $Q^r$, the stochastic equation for the angular velocity $\Omega$ of rotational motion reads
\begin{equation}
\begin{split}
I\frac {\mathrm{d}\Omega}{\mathrm{d}t}=Q^p+Q^f+Q^r \label{eqmotion}
\end{split}
\end{equation}
where $I$ is the moment of inertia of the rotor. The random torques are assumed to be white noises, having Gaussian distribution with $\overline{Q^r(t)}=0$ and $\overline{Q^r(t)Q^r(t+\tau)}=2D\delta(\tau)$. The noise strength $D$ will be determined in Sec. \uppercase\expandafter{\romannumeral4}.

The driving torque rotates the micro-rotor and concomitantly the thrust force of liquid sets up the translational motion.  The velocity field of the incompressible viscous liquid satisfies the Navier-Stokes equation
\begin{equation}
\begin{split}
\rho \frac{\partial \mathbf{V}}{\partial t}+\rho (\mathbf{V}\cdot\nabla)\mathbf{V}=-\nabla p+\eta\nabla^2\mathbf{V}
\label{NSE}
\end{split}
\end{equation}
and is subjected to the continuity equation
\begin{equation}
\begin{split}
\nabla \cdot \mathbf{V}=0 \label{cont}
\end{split}
\end{equation}
 where $\mathbf{V}$ represents the flow velocity, $\rho$ is the density of liquid, $p$ the pressure of liquid and $\eta$ the viscosity that has value $1.\mu g (\mu m \cdot s)^{-1}$ for water. The Reynolds number of the rotor with radius $r$ is
 \begin{equation}
 R=\frac{\rho |\Omega|r^2}{\eta} \label{reynolds}
 \end{equation}

 It is well known that the Navier-Stokes equation is simplified to the linear Stokes equation by dropping the nonlinear term of (\ref{NSE}) when the  Reynolds number $R$ is small. On the other hand, the viscosity can be ignored when $R$ is much larger than one.

\section{Driving torque and friction torque}

\subsection{Driving torque}

The average driving torque on unit volume of blades during a period of light, produced by the circularly polarized laser, is given by \cite{prost1995physics}
\begin{equation}
\langle \mathbf{P}(t)\times\mathbf{E}(t)\rangle
-\mathbf{r}\times \frac{\delta \langle \mathbf{E}(t)\cdot \mathbf{P}(t)\rangle}{\delta \mathbf{r}}
\label{qd}
\end{equation}
where $\langle \cdot \rangle$ denotes the time average, $\mathbf{E}(t)$ is the
electric field and $\mathbf{P}(t)$ is the electric polarization at the moment $t$. The dielectric response is characterized by the decay function $f_\mu(\tau)$ \cite{jackson1999classical}, i.e.
\begin{equation}
\begin{split}
P_\mu(t)=\epsilon_0\int_0^\infty f_\mu(\tau)E_\mu(t-\tau)\mathrm{d}\tau
\end{split}
\end{equation}
where $\mu=r,n$, and $t$ denote the components along three orthogonal unit vectors, namely, radial vector $\mathbf{e}_r$, vector perpendicular to the blade $\mathbf{e}_n$, and $\mathbf{e}_t=\mathbf{e}_n\times \mathbf{e}_r$. The complex permittivity $\epsilon_\mu(\omega)$ is related with $f_\mu$ by
\begin{equation}
\epsilon_\mu (\omega)=\epsilon_0+\epsilon_0\int_0^\infty e^{i\omega\tau}f_\mu(\tau)\mathrm{d}\tau
\end{equation}

We concentrate on the torque along $z$-axis, exerting on area $dA$ of the blade surface. From (\ref{qd}), it reads
\begin{equation}
\begin{split}
&\mathrm{d}Q^p=2a\epsilon_0\mathrm{d}A\frac{\delta}{\delta \varphi}\langle \int_0^\infty f_\mu(\tau)E_\mu (t-\tau)E_\mu (t)\mathrm{d}\tau\rangle_{r,z}\\
&+a\epsilon_0\mathrm{d}A\langle\int_0^\infty f_\mu E_\mu(t-\tau)[\mathrm{e}_{\mu x}E_y(t)-\mathrm{e}_{\mu y}E_x(t)]\rangle\\
\end{split} \label{Qpint}
\end{equation}
The projections of $\mathbf{e}_\mu$ in the $x$ and $y$ directions have been denoted by $\mathrm{e}_{\mu x}$ and $\mathrm{e}_{\mu y}$, respectively.

The circularly polarized laser has electric field $E_j=\mathrm{Re}[u_j(z,t)\exp(ikz-i\omega t+i\phi_j)]$ with $j=x,y$. For simplicity $u_j(z,t)$ is set to be a constant $u_0$, and the phase difference $\Delta \phi=\phi_y-\phi_x$ is $\sigma\pi/2$, with $\sigma=+1$ and $-1$ for left-handed and right-handed circularly polarizations respectively. Integration of equation $(\ref{Qpint})$ vanishes the first term, leading to
\begin{equation}
Q^p=3\sigma aSu_0^2  (\epsilon''_n(\omega)\cos^2\beta+\epsilon''_t(\omega)\sin^2\beta+\epsilon''_r(\omega)) \label{qp}
\end{equation}

The dielectric loss is characterized by the imaginary part of permittivity $\epsilon''$, which is responsible to the opto driving torque. Notably, even isotropy dielectric has non-zero averaged total opto driving torque along $z$-axis if $\epsilon''$ is not vanishing. It can be explained as the effect of angular momentum transferring from photon spin to the dielectric. The driving torque is proportional to the field intensity $u_0^2$ and total surface area $3S$ of the blades with direction depends on the rotating direction of the circularly polarized light as indicated by the sign of $\sigma$ in (\ref{qp}).

\subsection{Friction torque}

The friction torque on the rotor is given by
\begin{equation}
Q^f=|\mathbf{e}_z\cdot\int_A \mathbf{r}\times \mathbf{T} dA|
\end{equation}
where $\mathbf{T}$ is the total stress, deriving from the velocity gradient and the pressure of the fluid.

It is very difficult if not impossible to obtain $Q^f$ analytically because the complicated boundary. Before getting to the results of numerical simulation in Sec.\uppercase\expandafter{\romannumeral5}, let us discuss the scaling behaviors in the small and large Reynolds number regimes.

The friction torque acts as a resistance to the rotational motion of the rotor. One can express the friction torque as
\begin{equation}
Q^f = -\frac{f(\sqrt{R})}{\pi}\eta r^{3}\Omega  \label{qfn}
\end{equation}
where $f$ is a dimensionless constant for small Reynolds numbers or a function of $\sqrt{R}$ that covers the solution for the laminar boundary layer .\cite{landau1987fluid}  Expanding $f$ in powers of $\sqrt{R}$ one asymptotically has
\begin{equation}
f=c_{0}+c_{1}\sqrt{R}+c_{2}R+\dots \label{alpha}
\end{equation}
where $c_{n}$ ($n=0,1,2,\dots$) are dimensionless.
For small Reynolds numbers, the constant term of (\ref{alpha}) is dominated. Hence the friction torque exhibits a cubic power dependence on $r$ as
 \begin{equation}
 Q^f=-\frac{c_{0}}{\pi}\eta r^3 \Omega \label{qfs}
 \end{equation}
which is in consistent with the linear Stokes equation in the small Reynolds number regime.

For large $r$ but slow rotation, the edge effect is less important and the flow near to the surface of blades can be considered as a two-dimensional laminar boundary layer flow. In this case the second term of (\ref{alpha}) is dominated and the friction torque may be approximated by $Q^f=-\frac{c_{1}}{\pi}\sqrt{\rho \eta\Omega}r^{4}\Omega$. When both $r$ and the angular velocity of rotation are large, the third term of (\ref{alpha}) may enter and one has no simple scaling on $r$ but
\begin{equation}
Q^f=-\frac{c_{1}}{\pi}\sqrt{\rho \eta\Omega}r^{4}\Omega-\frac{c_{2}}{\pi}\rho r^{5}\Omega^2 \label{qfl}
\end{equation}
The last term does not depend on the viscosity, representing the drag torque of ideal flow (with turbulent in the boundary layer) that generally appears in the regime of large Reynolds numbers. Eq.(\ref{qfl}) is beyond the linear regime of the Navier-Stokes equation.

The coefficients $c_{n}$ ($n=0,1,2$) depend on the structure, for example the tilt angle $\beta$, and the boundary conditions. Generally they can only be obtained by numerical calculation. But for a thin disk plate with large radius thus the edge effect is negligible there is an analytical solution $Q^f=-1.94\sqrt{\rho\eta \Omega}r^4\Omega$. \cite{doi:10.1002/zamm.19210010401,landau1987fluid} The solution of thin disk plate will be used for inspection of our numerical simulation in Sec. \uppercase\expandafter{\romannumeral5}.

\section{Angular Velocity}

In this section we concentrate on the linear regime of the Navier-Stokes equation. Adopting (\ref{qfs}) for $Q^f$, and introducing $\alpha=\frac{c_{0}}{\pi}\eta r^3$,  the equation (\ref{eqmotion}) becomes
\begin{equation}
\begin{split}
I\frac {\mathrm{d}\Omega}{\mathrm{d}t}=Q^p-\alpha \Omega+Q^r \label{leqmotion}
\end{split}
\end{equation}

Solving (\ref{leqmotion}) and averaging the white random torques, one obtains
\begin{equation}
\overline{\Delta\Omega(t)^2}\equiv\overline{\Omega(t)^2}-\overline{\Omega(t)}^2=\frac{D}{I\alpha}(1-\exp^{-\frac{2\alpha}{I}t})
\end{equation}
In the stationary state ($t\gg \frac{I}{\alpha}$), $\overline{\Delta\Omega^2}=\frac{D}{I\alpha}$. According to the equipartition law of classical statistical mechanics $\overline{\Delta\Omega^2}=\frac{kT}{I}$, hence
\begin{equation}
D=kT\alpha
\end{equation}

The stochastic process can be described by the time-depending probability distribution of the angular velocity which satisfies the Fokker-Planck equation. The Fokker-Planck equation related to (\ref{leqmotion}) can be derived as
\begin{equation}
\begin{split}
&\frac{\partial}{\partial t} W(\Omega,t)\\&=-\frac{1}{I}\frac{\partial}{\partial \Omega} [(-\alpha\Omega+Q^p)W(\Omega,t)]+\frac{D}{I^2}   \frac{\partial^2}{\partial \Omega^2} W(\Omega,t)
\end{split}
\end{equation}
The stationary solution with the boundary condition $W(+\infty)=W(-\infty)=0$ is obtained as
\begin{equation}
\begin{split}
W(\Omega)=\sqrt{\frac{\alpha I}{2\pi D}}\exp[-\frac{\alpha I}{2 D}(\Omega-\frac{Q^p}{\alpha})^2]
\end{split}
\end{equation}

The general non-stationary solution is obtained by the Fourier transformation and the method of characteristics. Defining $\Omega'=\Omega-\frac{Q^p}{\alpha}$ and $W'(\Omega',t)=W(\Omega,t)$, equation(15) is rewritten as the Ornstein-Uhlenbeck equation\cite{Risken1996},
\begin{equation}
\begin{split}
&\frac{\partial}{\partial t} W'(\Omega',t)\\&=\frac{\alpha}{I}\frac{\partial}{\partial \Omega'} (\Omega' W'(\Omega,t))+\frac{D}{I^2}   \frac{\partial^2}{\partial \Omega'^2} W'(\Omega',t)
\end{split}
\end{equation}
Applying the initial condition $W(\Omega,0)=\delta(\Omega-0)$, i.e. $W'(\Omega',0)=\delta(\Omega'+\frac{Q^p}{\alpha})$, the non-stationary solution is obtained as
\begin{equation}
\begin{split}
W(\Omega,t)=&\sqrt{\frac{\alpha I}{2\pi D(1-\exp^{-\frac{2\alpha}{I}t})}}\\
&\exp[-\frac{\alpha I(\Omega-\frac{Q^p}{\alpha}+\frac{Q^p}{\alpha}\exp^{-\frac{\alpha}{I}t})^2}{2D(1-\exp^{-\frac{2\alpha}{I}t})}]
\end{split}
\end{equation}
Defining a time-dependent torque as
\begin{equation}
Q^p(t) =Q^p(1-\exp^{-\frac{\alpha}{I}t})
\end{equation}
and a time-dependent diffusion coefficient as,
\begin{equation}
D(t)=D(1-\exp^{-\frac{2\alpha}{I}t})
\end{equation}
both the stationary and the non-stationary solution can be written in the same form as
\begin{equation}
\begin{split}
W(\Omega,t)=\sqrt{\frac{\alpha I}{2\pi D(t)}}\exp[-\frac{\alpha I}{2 D(t)}(\Omega-\frac{Q^p(t)}{\alpha})^2]
\end{split} \label{Wt}
\end{equation}
Obviously, when $t\rightarrow\infty$ it approaches to the stationary solution.

At each moment $t$, $\Omega$ has a Gaussian distribution with the mean
\begin{equation}
\Omega_0(t) =\frac{Q^p}{\alpha}(1-\exp^{-\frac{\alpha}{I}t})
\end{equation}
and the standard deviation
\begin{equation}
\begin{split}
\sigma_\Omega(t) &=\sqrt{\frac{D}{\alpha I}(1-\exp^{-\frac{2\alpha}{I}t})}\\
&=\sqrt{\frac{kT}{I}(1-\exp^{-\frac{2\alpha}{I}t})}
\end{split}
\end{equation}
As the time increases, both $\Omega_0$ and $\sigma_\Omega$ increase and reach their stationary values at $t\gg \frac{I}{\alpha}$ and $t\gg \frac{I}{2\alpha}$, respectively. For larger $Q^p$ or smaller $\alpha$, $\Omega_0(\infty)$ has larger value. Namely, higher intensity of the driving laser, larger imaginary permittivity of the blade material and lower liquid viscosity all lead to higher stationary angular velocity. On the other hand, the fluctuation is proportional to $\sqrt{\frac{kT}{I}}$. Hence, at a given temperature, the fluctuation is important for micro-rotors which have small inertia moment. Moreover, the hollow CMT\cite{WEN20114067} is allowed to load, which increases $I$, thereby slows the process and suppresses the fluctuation, but does not change the final mean velocity.

\section{simulation}

The rotor is propelled forward in the liquid by the laser beams. We adopt the inertial frame of reference in which the rotor has no translational motion, while the flow at infinity has a constant axial velocity. In simulation, the rotor is placed at the center of a water-filled cylinder vessel with the radius of $12r$ and the height of $42h$. The fluid velocity on the top and bottom surfaces of the vessel satisfies the periodic boundary condition, while the side surface of the vessel is moving with a constant velocity $-V_z$. Iterate $-V_z$ and the velocity of the top (or bottom) surface until that the difference between them is minimized. $V_z$ is the estimated translational velocity of the rotor while it rotates with the given angular velocity in the laboratory frame of reference. The steady-state solver for the Navier-Stokes equation (\ref{NSE}) subjected to (\ref{cont}) has the relative tolerance within $0.001$.

\begin{figure}
  \centering
  \subfigure[]{\includegraphics[width=8.6cm]{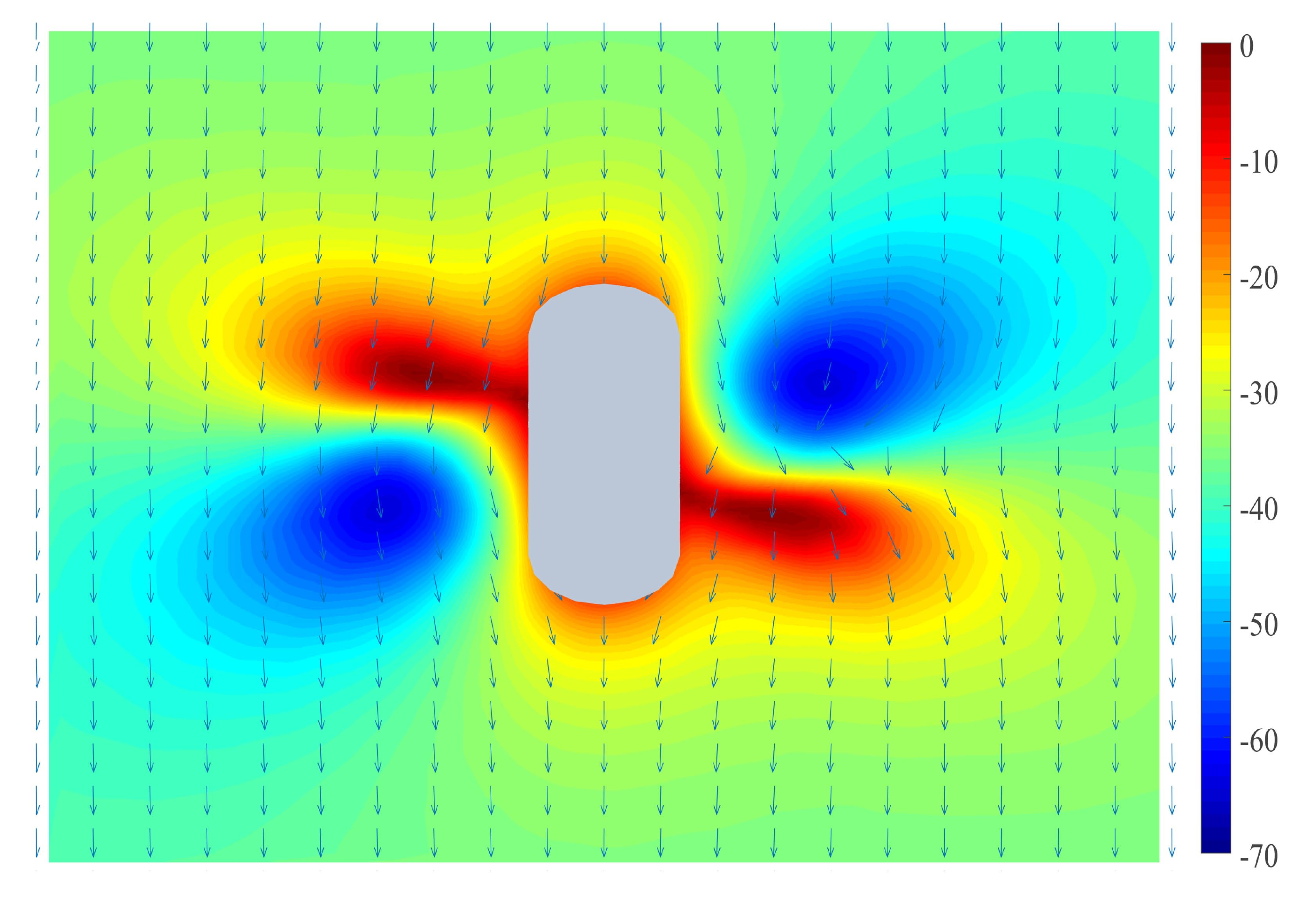}}
  \subfigure[]{\includegraphics[width=8.6cm]{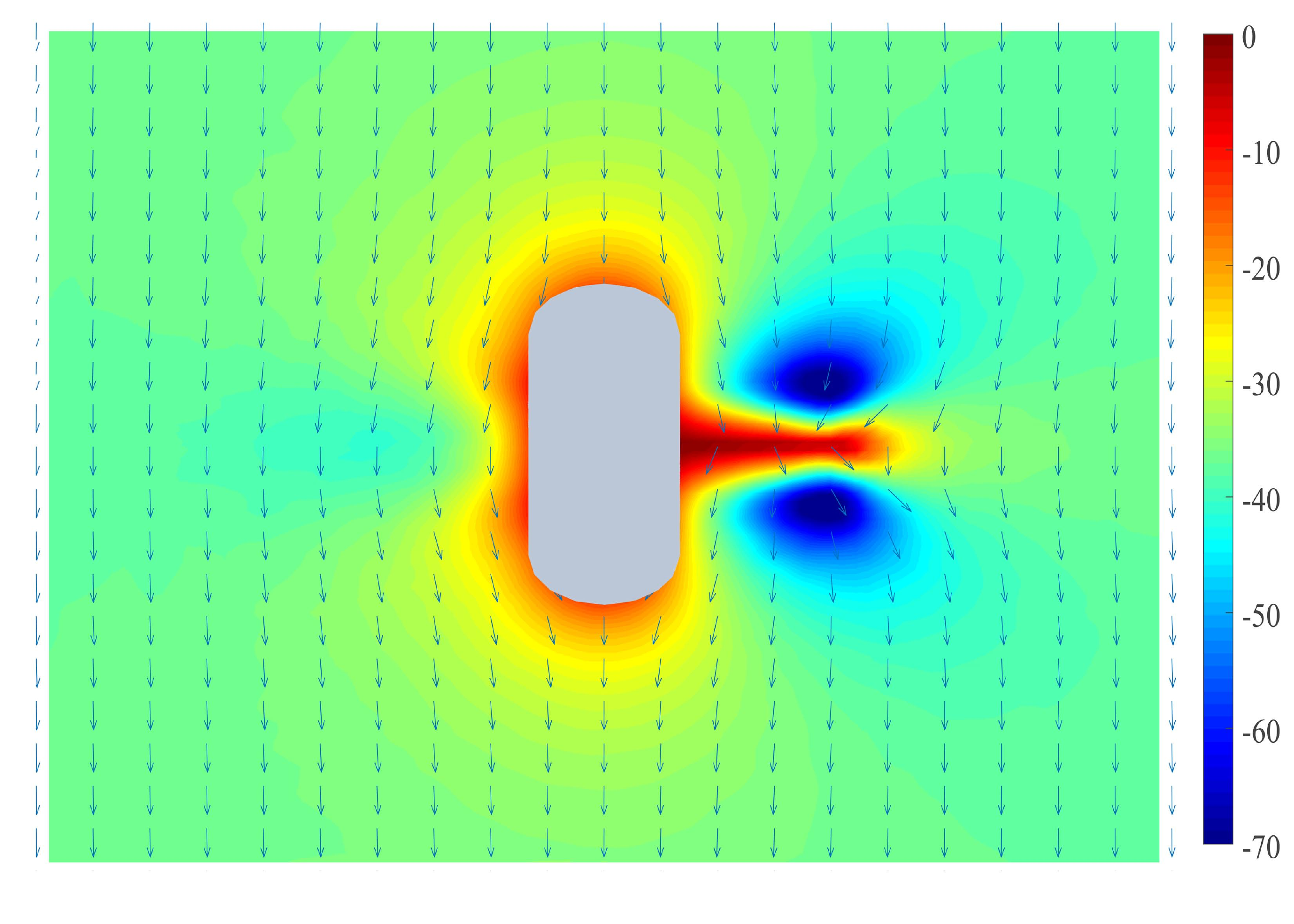}}\\
  \caption{The $z$-component of fluid velocity in (a) the $yz$-plane and (b) the $xz$-plane represented by color map. The arrows indicate the projected directions of the velocity field on the corresponding planes. The rotor has $r=2.10\mu m$, $\Omega=60\pi s^{-1}$ and $\beta=32^\circ$.}\label{Fig.v}
\end{figure}

\begin{figure}
  \centering
  \includegraphics[width=8.6cm]{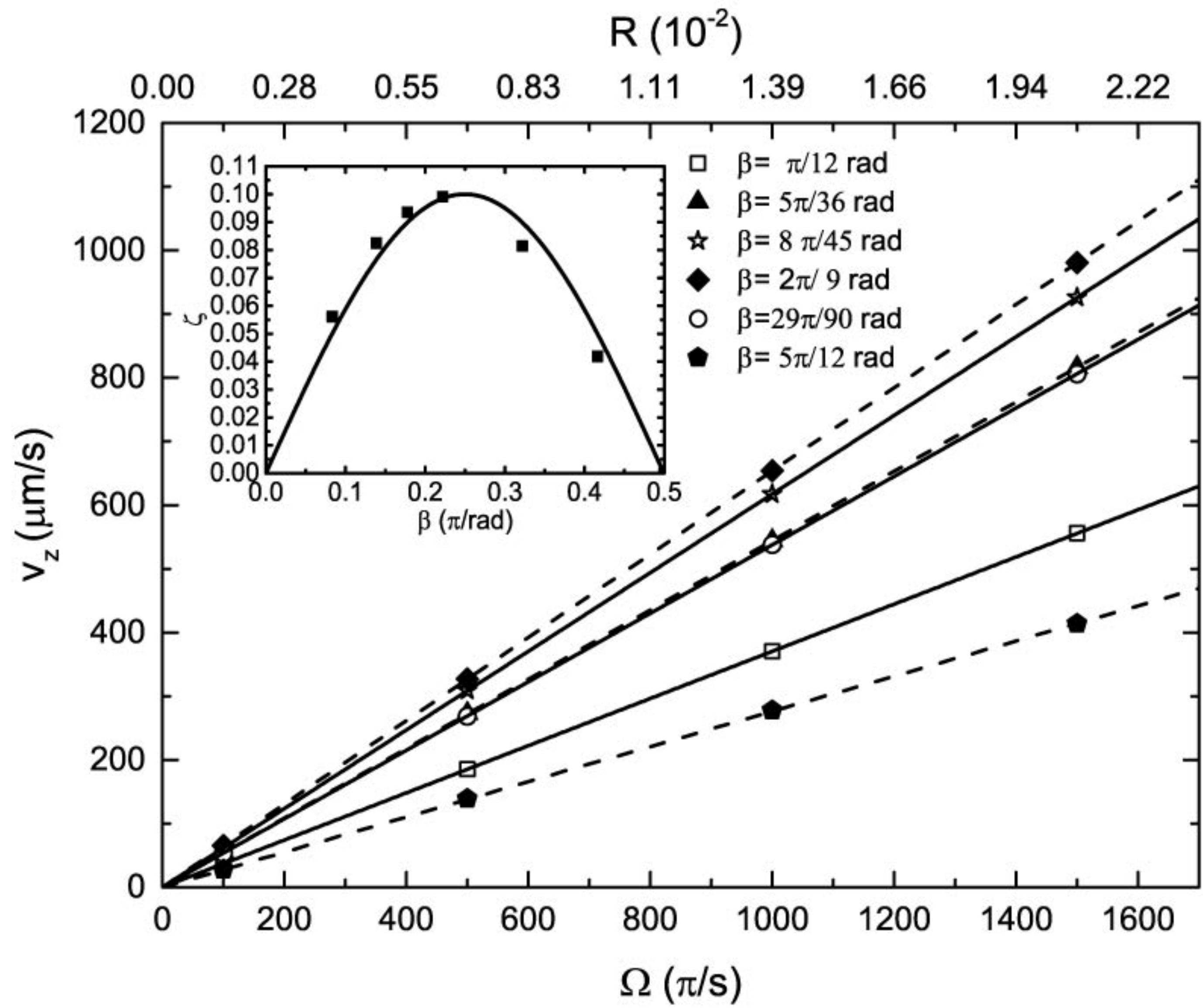}\\
  \caption{The translational velocity $V_z$ of the micro-rotor having $r=2.10\mu m$, which exhibits linear relation with angular velocity $\Omega$ with the slope depending on the tilt angle $\beta$. The inset shows the propelling efficiency $\zeta$ versus $\beta$.  The top axis is the Reynolds number with unit $10^{-2}$.}\label{Fig.Vzwbeta}
\end{figure}

We first simulate a representative micro-rotor whose axle has radius of $r_0=0.45\mu m$ and height of $h=2.90\mu m$, and each blade has radius of $r=2.10\mu m$, area of $S=1.65\mu m\times 1.33 \mu m$ and thickness of $a=0.021\mu m$. Fig.\ref{Fig.v} (a) and (b) display the axial velocity (color map) for the tilt angle $\beta=32^\circ$ and $\Omega=60\pi s^{-1}$ in the $yz$-plane and the $xz$-plane respectively. The projected directions of the velocity field on the two planes are indicated by arrows.

\begin{figure}
  \centering
  \includegraphics[width=8.6cm]{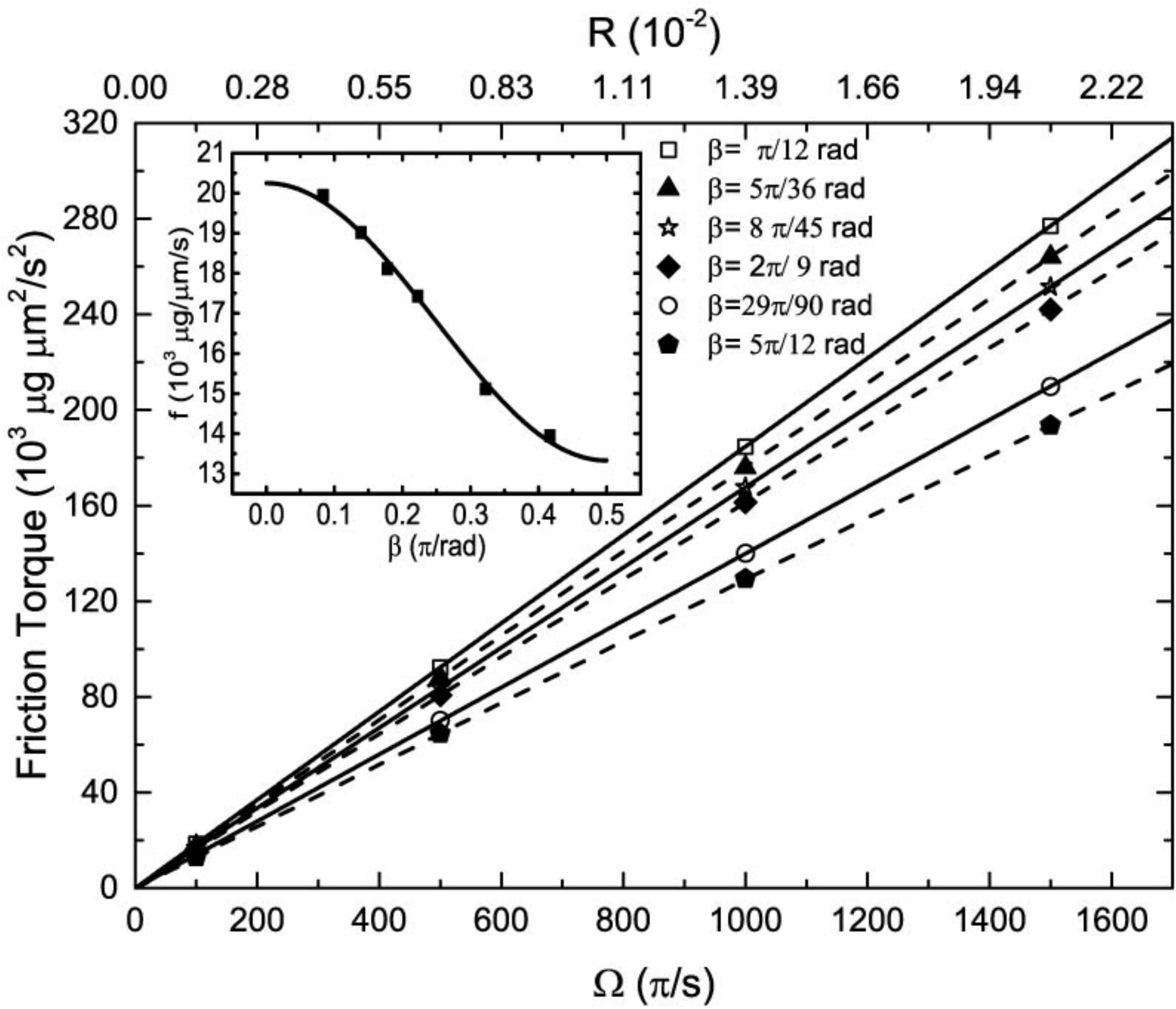}\\
  \caption{The absolute values of the total friction torque $Q^f$ of the micro-rotor having $r=2.10\mu m$, which exhibits linear relation with angular velocity $\Omega$ with the slope depending on the tilt angle $\beta$. The inset shows $f(\beta)$ versus $\beta$. The top axis is the Reynolds number with unit $10^{-2}$}\label{Fig.Qfwbeta}
\end{figure}

\begin{figure}
  \centering
  \includegraphics[width=8.6cm]{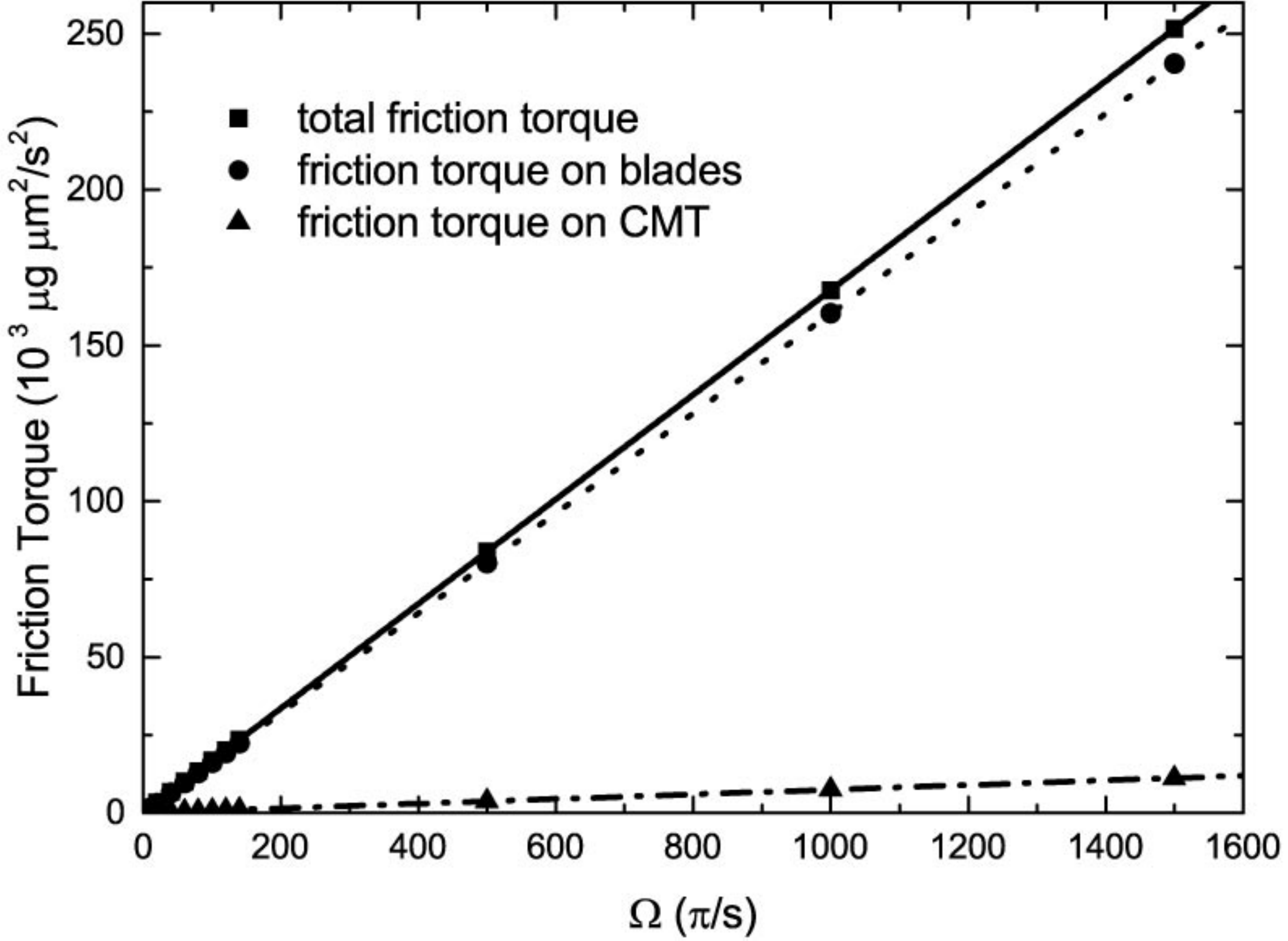}\\
  \caption{The linear increasing relationship of the absolute values of the total friction torque $Q^f$, the friction torque on blades, and the friction torque on CMT with the angular velocity $\Omega$. The friction torque on blades is larger than the friction torque on CMT by one or two orders of magnitude. The micro-rotor has $r=2.10\mu m$ and $\beta=32^\circ$.}\label{Fig.torque}
\end{figure}

Then rotors with various tilt angles $\beta$ and radii $r=0.21, 2.10, 21.0\mu m$ are simulated with the other parameters of length having the same ratios to $r$ as the representative micro-rotor. In the explored range of $R<2.$, the translational velocities and the friction torques for various tilt angles and radii are confirmed as linear functions of the angular velocity. The translational velocity and friction torque of $r=2.10\mu m$ are shown in (Fig.\ref{Fig.Vzwbeta}) and (Fig.\ref{Fig.Qfwbeta}) respectively. The propelling efficiency $\zeta\equiv\frac{V_z}{r\Omega}$ and the dimensionless quantity $f \equiv -\frac{\pi Q^f}{\eta r^{3}\Omega}$ are almost independent of $r$. The inset of Fig.\ref{Fig.Vzwbeta} shows the propelling efficiency $\zeta$ versus $\beta$, which can be approximated by $\zeta=0.1 \sin(2\beta)$ with the maximum at $\beta=\frac{\pi}{4}$. The dimensionless quantity $f\sim c_{0}$ in this regime can be approximated by $c_{0}=20.-6.9\sin^{2}\beta$ as shown in the inset of Fig.\ref{Fig.Qfwbeta}(solid line). In addition, the friction torque for blades is larger than the friction torque for CMT by one or two orders of magnitude (Fig.\ref{Fig.torque}).

\begin{figure}
  \centering
  \includegraphics[width=8.6cm]{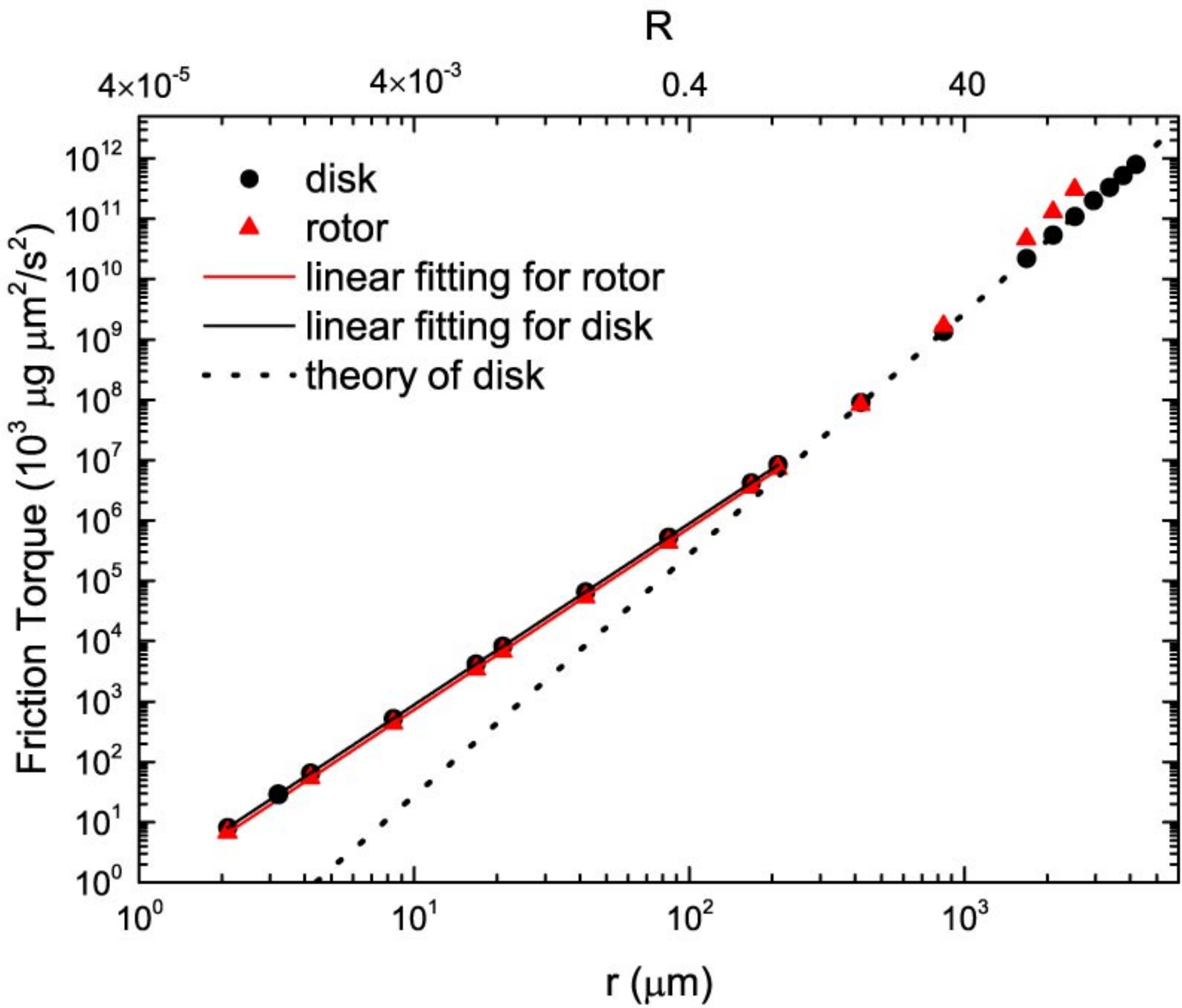}\\
  \caption{The friction torque versus radius $r$ (bottom axis) and the Reynolds number (top axis) at $\Omega=40\pi s^{-1}$ in log-scales. The triangles are simulation results for rotors of $\beta=32^\circ$ and the circles for the thin disks.  The triangles and circles in the regime of small $r$ (or $R$) are fitted by two solid lines respectively. Both solid lines have slope $3$, confirming the pow law $Q^f\propto r^3$ in the small Reynolds regime. The dotted line is the analytical solution for the thin disk model that is $Q^f=-1.94\sqrt{\rho\eta \Omega}r^4\Omega$.}\label{Fig.scale1}
\end{figure}

 To explore the scaling crossover from small to large Reynolds numbers, rotors of $\beta=32^{\circ}$ and thin disks with $r$ ranged from $2.1\mu m$ to $4. mm$ are simulated with angular velocity fixed at $\Omega=40\pi s^{-1}$. The friction torques versus $r$ and $R$ in log-scales are given in Fig.\ref{Fig.scale1}, where red triangles are for rotors and dark circles for thin disks. It shows clearly that the small Reynolds regime and large Reynolds regime have distinguished scaling laws. The numerical results for both micro-rotor and thin disk can be perfectly fitted by $Q^f\propto r^{3}$ for $r<200\mu m$, but exhibit obvious deviation for $r>200\mu m$ where the Reynolds number is much larger than one. The $r^3$ scaling is in consistent with the dimensional argument of  Sec.\uppercase\expandafter{\romannumeral3} on the friction torque for small Reynolds number. The results of the thin disk are well fitted by $\sim r^4$ in the large $r$ regime, consisting with the analytical solution for the thin disk of large radius.\cite{doi:10.1002/zamm.19210010401,landau1987fluid}

\section{Fluctuation and averaged velocity}

\begin{figure}
  \centering
  \includegraphics[width=8.6cm]{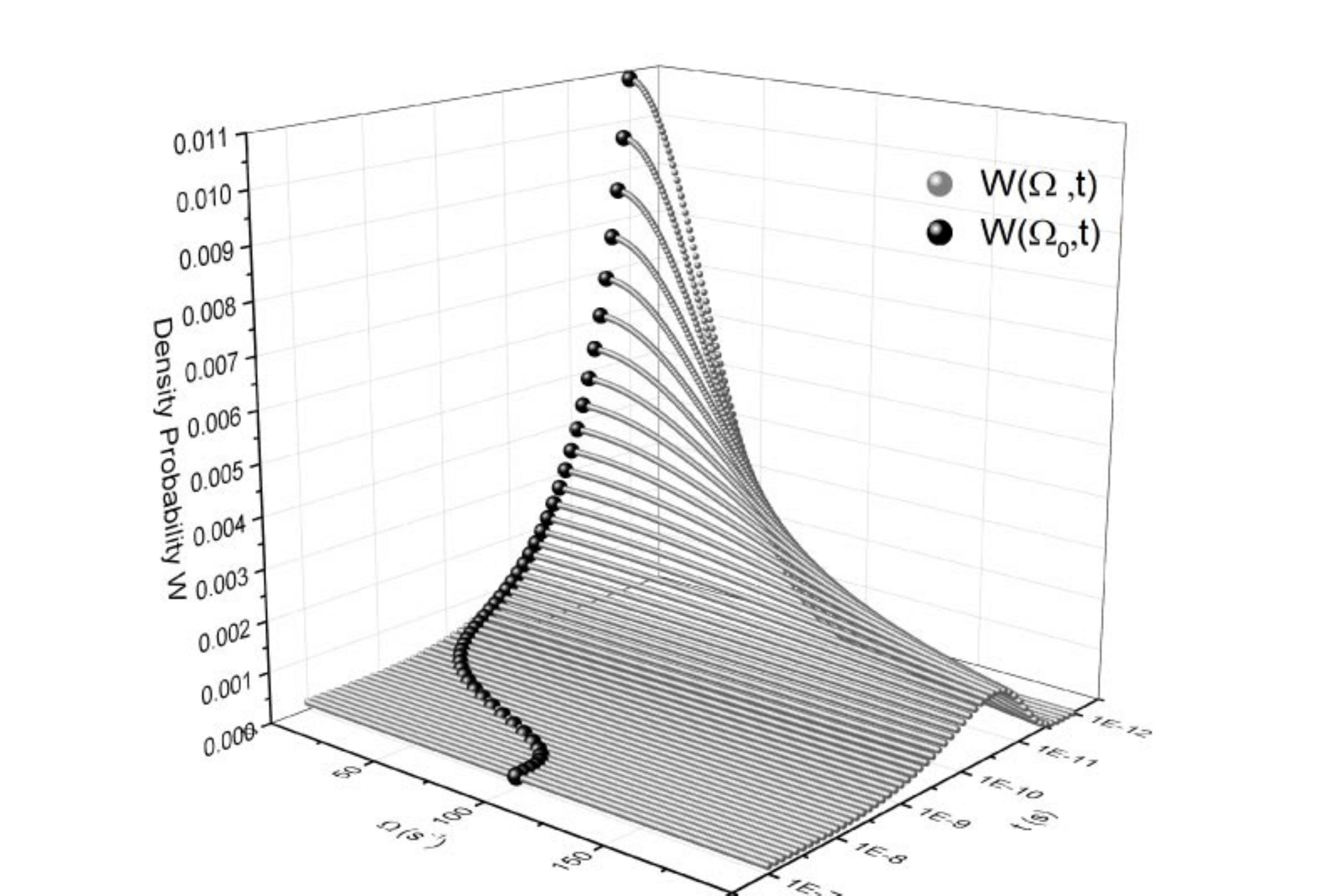}\\
  \caption{The time-dependent probability density $W(\Omega,t)$. The maximum at a given moment $W(\Omega_0,t)$
   is marked by a black ball. The result is computed with $r=2.10\mu m$, $\beta=32^\circ$, and $I=3.81\times 10^{-7}\mu g\cdot \mu m^2$. }\label{Fig.Womegat1}
\end{figure}

\begin{figure}
  \centering
  \includegraphics[width=8.6cm]{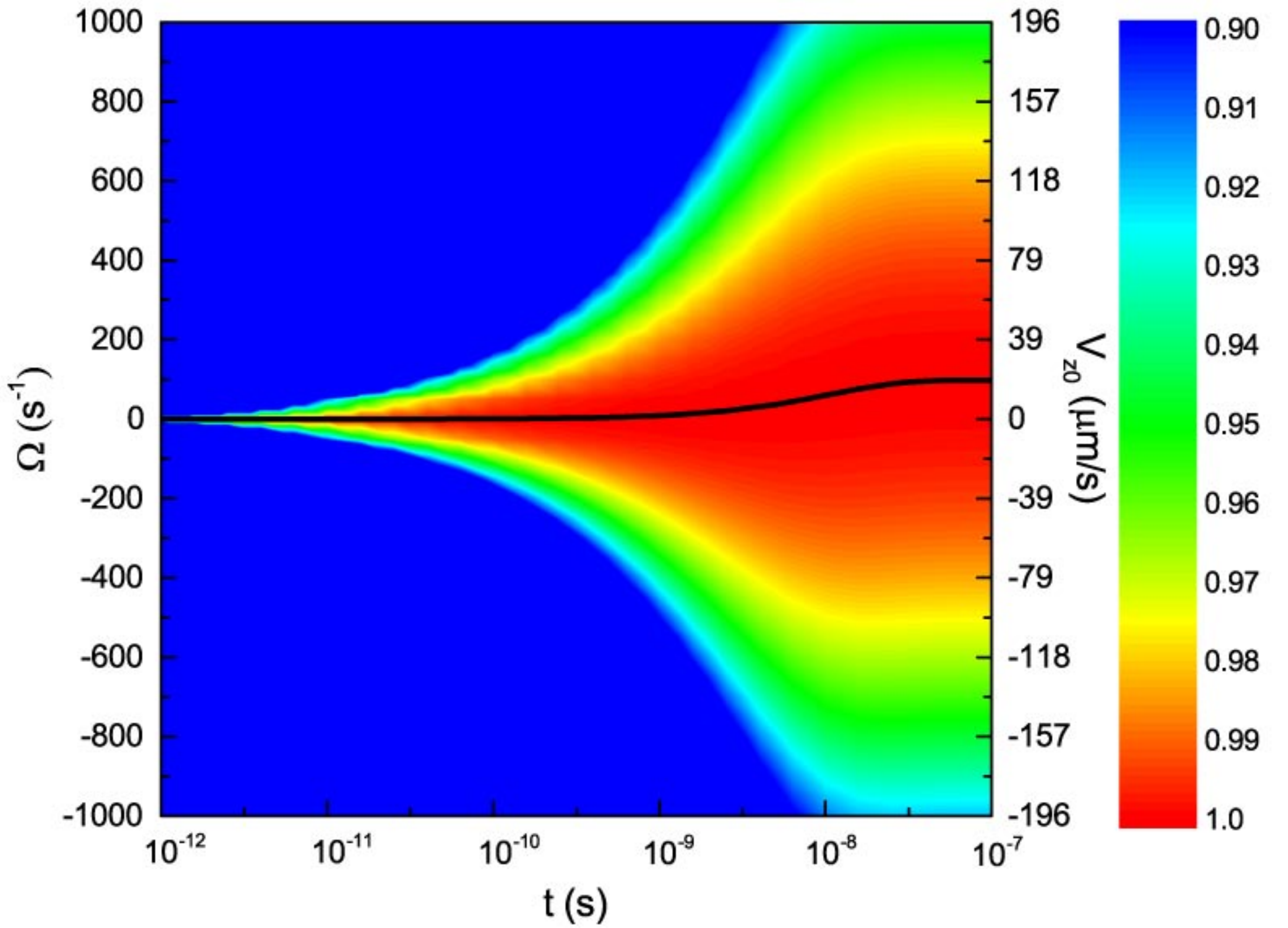}\\
  \caption{The time-depending normalized probability density $\tilde{W}(\Omega,t)$. The black solid line is the average angular velocity $\Omega_0(t)$. The density is cut off at $0.50$}\label{Fig.Womegat}
\end{figure}

\begin{figure}
  \centering
  \subfigure[]{\includegraphics[width=8.6cm]{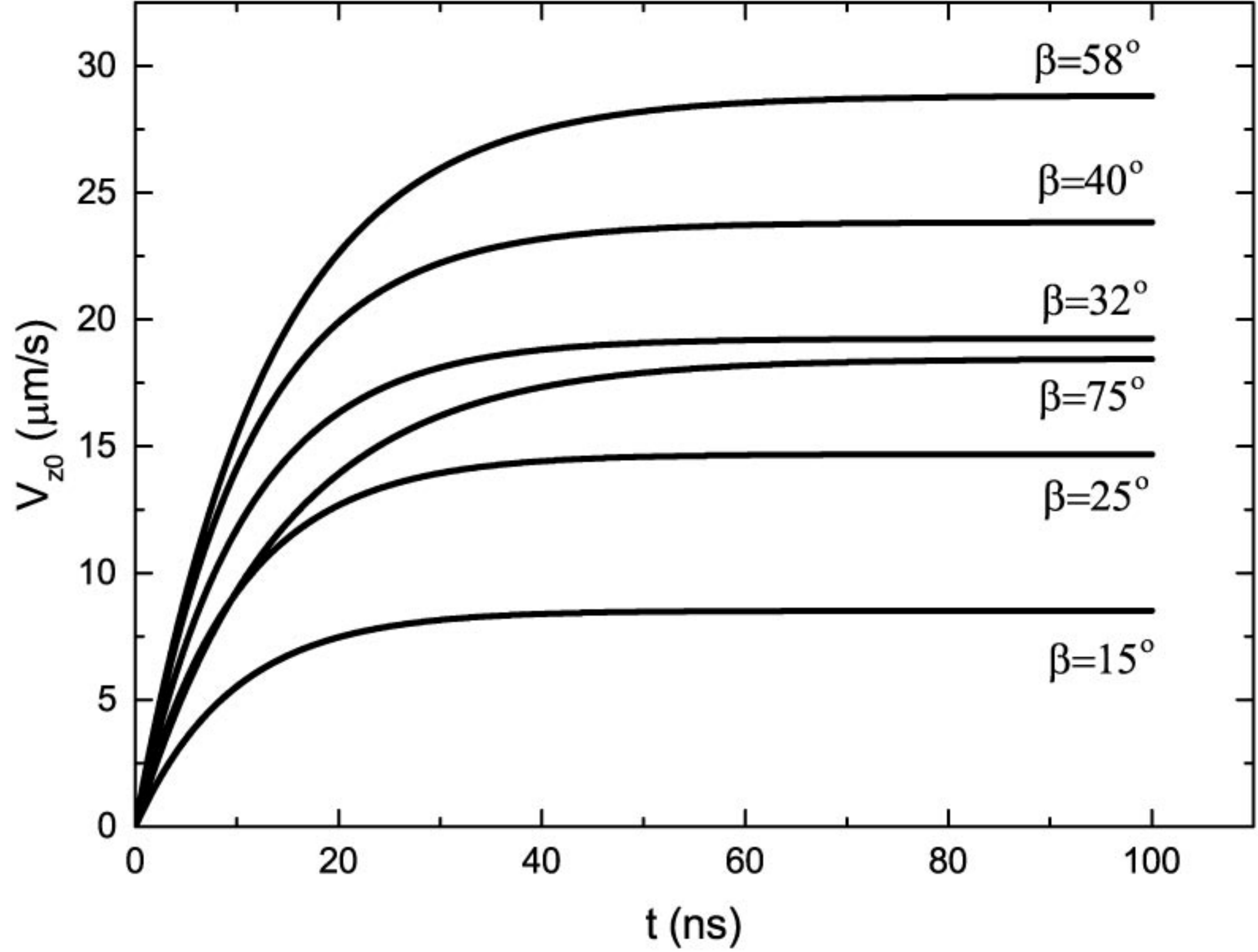}}
  \subfigure[]{\includegraphics[width=8.6cm]{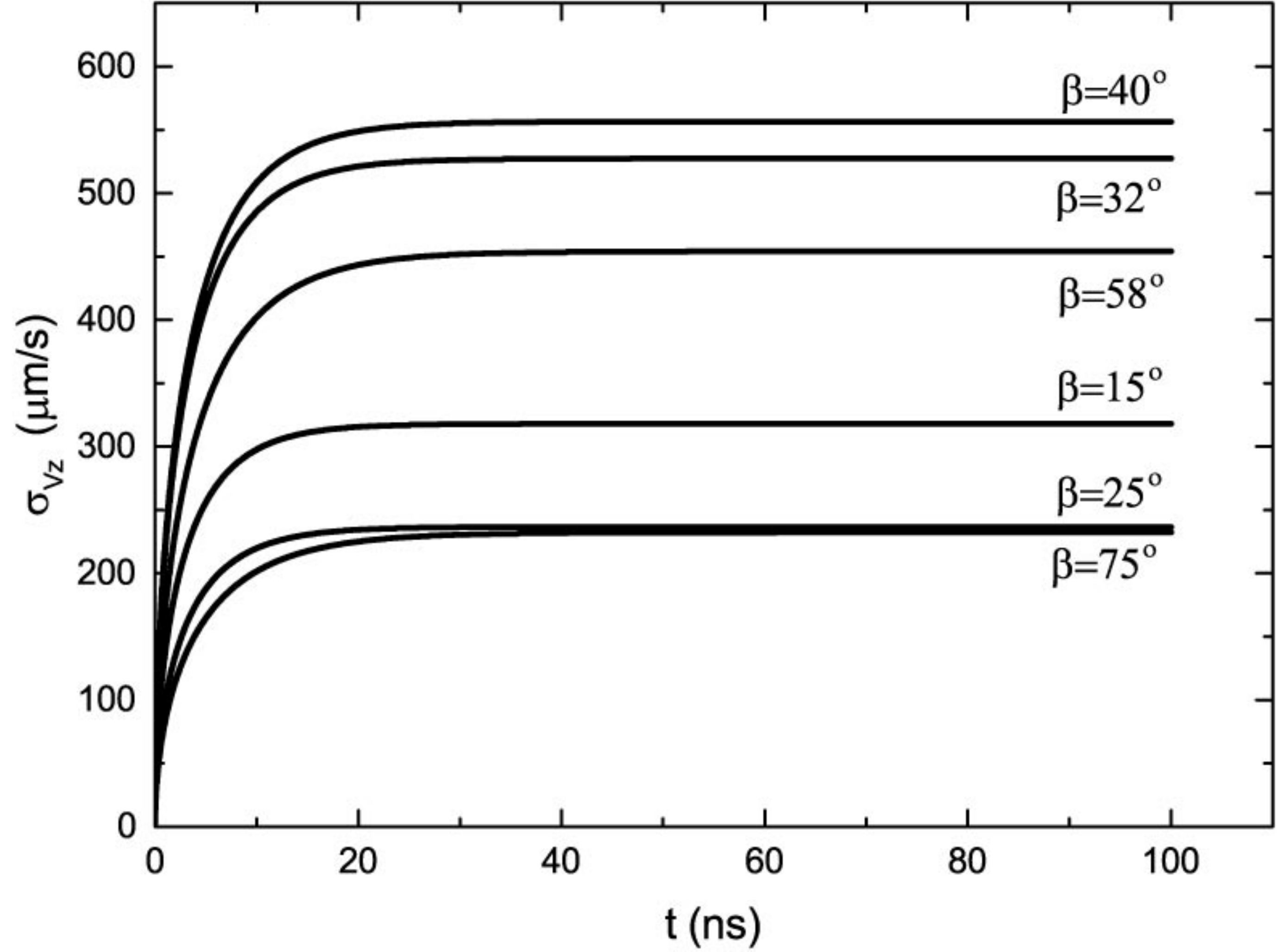}}\\
  \caption{(a) The time-depending mean translational velocity $V_{z0}(t)$ and (b) the standard deviation $\sigma_{V_z}(t)$.
  }\label{Fig.omegat}
\end{figure}

To be specific, the wave length of the driving laser is chosen to be $543nm$ and the amplitude of electric field $u_0=1V/\mu m$. The corresponding imaginary permittivity of the graphite \cite{doi:10.1002/sia.3522,doi:10.1063/1.3562033} $\epsilon''_n(\omega)\approx0$ and $\epsilon''_t(\omega)=\epsilon''_r(\omega)=3.23\epsilon_0$. From (\ref{qp}), the average opto driving torque for $r=2.1\mu m$ and $\beta=32^\circ$ is $|Q_z^p|=5.23\times 10^{3}\mu g\cdot\mu m^2/s^2$.

We again concentrate on the small Reynolds regime where the friction torque is linear with the angular velocity. Adopting the numerical $c_{0}$ and the corresponding $\alpha$ for the representative micro-rotor obtained in the simulation, we  plot the probability density $W(\Omega,t)$ of (\ref{Wt}) in Fig.\ref{Fig.Womegat1} for $I=5.67\times 10^{-7}\mu g\cdot\mu m^2$. Black balls are  $W(\Omega_0,t)$, the maximum probability density at each moment. To exhibit the time-dependent fluctuation of angular velocity more clearly, a normalized probability density is introduced as
\begin{equation}
\tilde{W}(\Omega,t)=W(\Omega,t)/W(\Omega_0,t)
\end{equation}
In Fig.\ref{Fig.Womegat}, color map of $\tilde{W}(\Omega,t)$ is presented, with cut-off at value of 0.9. The black solid line is corresponding to the time-depending mean angular velocity $\Omega_0(t)$. The normalized probability density $\tilde{W}(\Omega,t)$ has the maximum value ($1$) along the black solid line.

From Fig.\ref{Fig.omegat}, one sees that the mean translational velocity $V_{z0}$ and the standard deviation $\sigma_{V_z}(t)$ are increasing exponentially at the earlier stage and finally saturated at $t\gg \frac{I}{\alpha}$ and $t\gg \frac{I}{2\alpha}$, respectively.

\section{summary}

In summary, we investigate the hydrodynamic behavior of the micro-rotor model, which is rotated by two counter-propagating circularly polarized laser beams. The dielectric loss contributes to a finite and controllable driving torque that propels the micro-rotor to translational motion in the direction depending on the handedness. The friction torque and the propelling efficiency as two essential quantities of the problem are calculated. In the range of radius $r<200\mu m$ and the Reynolds numbers smaller than $5.$, it is found that the hydrodynamic friction torque is a linear function of the angular velocity with the slope proportional to $r^3$, which is in consistent with the dimension analysis on the linear Stokes theory and different from the $r^4$ scaling law predicted by the theory of laminar boundary layer. Therefore the $r^3$ scaling law in the regime of small Reynolds numbers as a consequence of the linear Stokes theory also implies the breakdown of the theory of laminar boundary layer and the necessity of numerical simulation for micro-systems whence the edge effect is significant. The consistence between the simulation and the dimension analysis as well as the exact solution for thin disk of large radius supports the validity of our simulations.

In the small Reynolds number regime, the stochastic fluctuation of the angular velocity as a function of time is obtained  as  the solution of the Fokker-Planck equation. The fluctuation is about an order higher than the mean value  for the representative micro-rotor. It can be reduced by increasing the moment of inertia.
The steady mean angular velocity is $\Omega_0(\infty)=\frac{\pi Q^p}{c_{0}\eta r^{3}}$, corresponding to the steady mean translational velocity $V_z=\frac{\pi \zeta(\beta)Q^p}{c_{0}\eta r^{2}}$ with $Q^p$ the opto driving torque. Numerically, the dimensionless coefficient $c_{0}$ depends on $\beta$ as $c_{0}=20.-6.9\sin^2 \beta$, where $\beta$ is tilt angle. The propelling efficiency can be approximated by $\zeta=0.1\sin(2\beta)$, which is almost independent of the radius of the rotor. The tilt angle $\frac{\pi}{4}$ maximizing the propelling efficiency is in accordance with the proportion of pitch and radius given by M.Doi and M.Makino\cite{doi:10.1063/1.2107867,doi:10.1063/1.4962411}.

Driving by circularly polarized laser with a feasible field strength of $1V/\mu m$, the micro-rotor can translate along the axial direction at a speed of twenty micrometers per second, enabling the handedness separation for the asymmetry microsystems.  Remarkably, filling load inside the CMT only delays the stationary motion but not reduces the final steady velocity. Therefore the micro-rotor we proposed can be used as a controllable carrier for delivering drugs or other tiny particles in liquid.

\begin{acknowledgments}
We thank C. Flytzanis for drawing our attention to this topic and having valuable discussions. The authors also thank the reviewers for suggesting the method of dimension analysis and for providing valuable references. The project is supported by the National Basic Research
Program of China (Grant: 2013CB933601), the National Key Research and Development Project of China
(Grant: 2016YFA0202001), and Special Program for Applied Research on Super Computation of the NSFC-Guangdong Joint Fund (the second phase).
\end{acknowledgments}

%%%%%%%%%%%%%%%%%%%%%%%%%%%%%%%%%%%%%%%%%%%%%%%%%%%%
%\bibliography{refphononrixs}
\bibliography{ref1}
\end{document}